\documentclass[twocolumn,aps,reprint,showpacs,preprintnumbers,amsmath,amssymb,groupedaddress]{revtex4-1}
\usepackage{graphicx}
\usepackage{dcolumn}
\usepackage{bm}
\usepackage{graphicx}
\usepackage{subfigure}
\usepackage{physics}

\begin{document}
\title{Thermoelectric cooling and thermal switching via the non-linear phonon Peltier effect}
	\author{Bitan De}
	\author{Bhaskaran Muralidharan}
	\email{bm@ee.iitb.ac.in}
	\affiliation{Department of Electrical Engineering, Indian Institute of Technology Bombay, Powai, Mumbai-400076, India}
	\date{\today}
\begin{abstract}   
		\indent Investigating the non-linear transport regime in a quantum dot heat engine described by the Anderson-Holstein model, it is shown that a finite electron-phonon interaction leads to a charge induced phonon generation that stimulates a phonon current even in the absence of a thermal gradient. This gives rise to  the non-linear phonon Peltier effect which shows a non-trivial dependence on varying the electron-phonon interaction. Utilizing the reversal of phonon currents via charge induced phonon accumulation, we demonstrate that the heat engine surprisingly can be cooled when coupled to a hot reservoir. In further exploring possibilities that can arise from this effect, we propose a charge-induced phonon switching mechanism.\\
	\end{abstract}
	\maketitle
\begin{figure}[]
		\begin{center}
			\includegraphics[width=0.45\textwidth, height=0.20\textwidth]{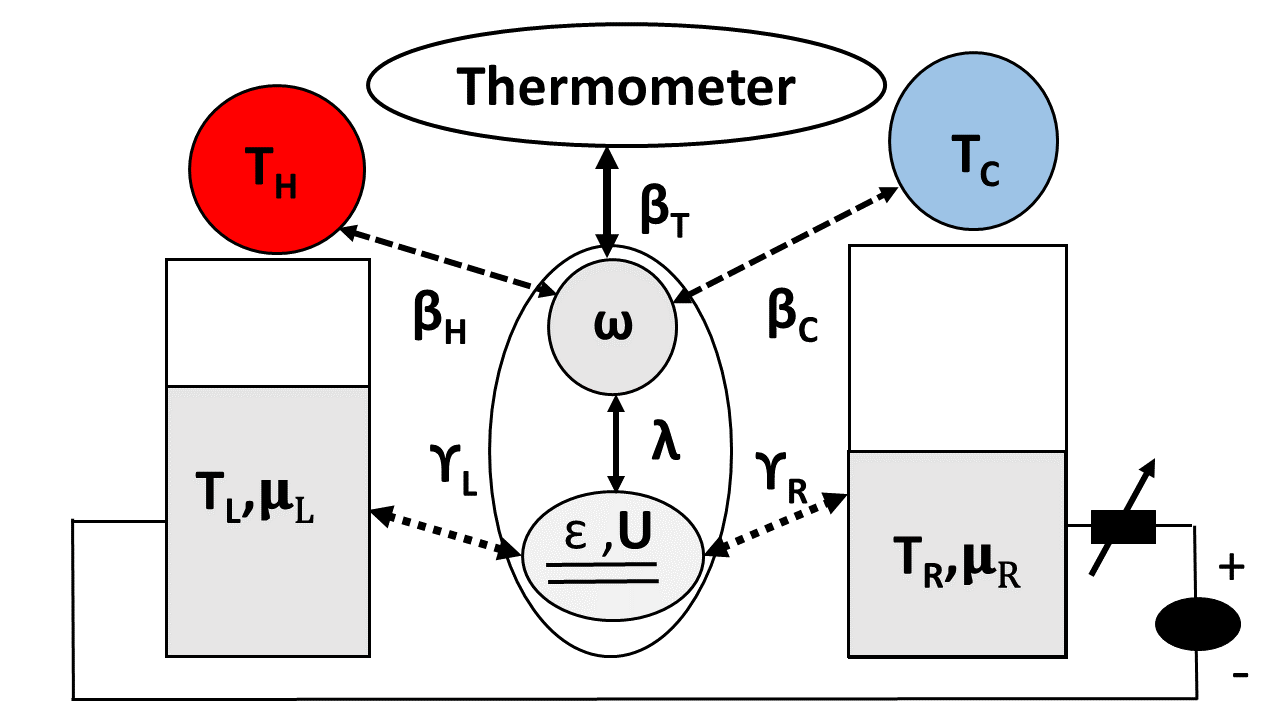}\label{q1}
		\end{center}
		\caption{Device Schematics: The quantum dot is weakly coupled to two electronic leads $L,R$ described by their respective chemical potentials and phonon reservoirs $H,C$ described by their respective temperatures . The corresponding electronic tunneling rate and phonon relaxation rate are $\gamma_{L(R)}$ and $\beta_{H(C)}$ respectively. Charge and phonon transport are coupled through electron-phonon interaction parameter $\lambda$, which drives the dot phonons out of equilibrium. The non-equlibrium phonon temperature is estimated using a thermometer bath.}
		\label{Fig:1}
	\end{figure}

{\it{Introduction.}} The study of quantum dot heat engines are invaluable when it comes to fundamental aspects of nanoscale heat flow. Significant interest is triggered from two principal motives: (a) Waste heat harvesting into electrical power \cite{Sothmann2014,Reddy1568,Kim2014,sc2016} and (b) Implementation of thermal energy in electronic logic design or 'phonon computation' \cite{Jiang,Bli1,Bli2,Karl,Bli3}.  A recent and notable work \cite{Ilani} exploited the nanoscale interaction of charge and vibrational degrees of freedom via the use of electrostatic gating to selectively couple charge transport with vibrational modes. This opens up new possibilities for optimizing thermo-electric performance \cite{de} as well as thermal switching linked to phonon computation.\\
\indent The most notable consequence of electron-phonon interaction is that the charge transport stimulated by an electrical voltage drives phonons away from equilibrium to initiate a phonon current even in the absence of any temperature gradient \cite{Leijnse2010,Siddiqui2006}. The dynamics of such a charge induced phonon generation can be explored using the voltage controlled dissipative heat engine set up \cite{Leijnse2010,Muralidharan2012,de} considered in Fig.~\ref{Fig:1}. Finite electron-phonon interaction modifies the phonon distribution and modulates the working temperature of the heat engine which can be recorded by a phonon thermometer weakly coupled to it \cite{Galperin2004,de}. In an earlier work \cite{de} we explored how charge-phonon interplay can influence the power-efficiency trade-off thereby influencing the thermoelectric operation. In this letter, we theoretically demonstrate how device cooling and thermal switching can be both accomplished by tailoring electron-phonon interactions.
In the context of cooling, most of the theoretical works \cite{oppen1,Adrian1} employ electronic heat currents across devices asymmetrically coupled to the contacts, while, our cooling proposal uses phonon heat currents in a symmetrically coupled device. The proposed switching mechanism here also concerns the modulation of charge currents via a thermal bias and the tailoring of the electron-phonon interaction.\\
\indent {\it{Set up and Formulation.}} The set up depicted in Fig.~\ref{Fig:1} comprises a quantum dot described by the dissipative \textit{Anderson-Holstein} model. Charge and phonon currents are set up by applying a voltage and thermal bias across the metallic electrodes $\alpha_1$ ($\alpha_1\in L,R$) and the thermal reservoirs $\alpha_2$($\alpha_2\in H,C$) respectively. \\
\indent The composite system Hamiltonian, $\hat{H}$, is the sum of the Hamiltonians of the dot ($\hat{H}_{D}$), the electrodes ($\hat{H}_{\alpha_1}$), the reservoirs ($\hat{H}_{\alpha_2}$) and the tunneling part ($\hat{H}_{T}$). The dot Hamiltonian is described as $\hat{H}_{D}=\hat{H}_D^{el}+\hat{H}_D^{ph}+\hat{H}_D^{el-ph}$ where $\hat{H}_{D}^{el}$, $\hat{H}_D^{ph}$, and $\hat{H}_D^{el-ph}$ represent the electronic part, the phonon part and the electron-phonon interaction part respectively. The electronic part $\hat{H}_{D}^{el}=(\sum_{\sigma}^{}\epsilon_{\sigma}\hat{d}_{\sigma}^{\dagger}\hat{d}_{\sigma}+U\hat{d}_{\uparrow}^{\dagger}\hat{d}_{\uparrow}\hat{d}_{\downarrow}^{\dagger}\hat{d}_{\downarrow})$, describes the dot in terms of a single spin degenerate energy level with energy $\epsilon_{\sigma}$ ($\sigma\in \uparrow,\downarrow$) and Coulomb interaction energy $U$. The phonon part $\hat{H}_{D}^{ph}=\hbar\omega_{\nu}\hat{b}_{\nu}^{\dagger}\hat{b}_{\nu}$ represents a single phonon mode with frequency $\omega_{\nu}$. In the expressions for $\hat{H}_{D}^{el}$ and $\hat{H}_D^{ph}$ , $\hat{d}_{\sigma}^{\dagger} (\hat{d}_{\sigma})$  and $\hat{b}_{\nu}^{\dagger} (\hat{b}_{\nu})$ denote the creation (annihilation) operator of the dot electrons and the dot phonons  respectively. In the dot, the electrons and phonons interact through a dimensionless coupling parameter, $\lambda_{\nu}$, and the corresponding interaction Hamiltonian is represented as $\hat{H}_{D}^{el-ph}=\sum_{\sigma}^{}\lambda_{\nu}\hbar\omega_{\nu}\hat{d}_{\sigma}^{\dagger}\hat{d}_{\sigma}(\hat{b}_{\nu}^{\dagger}+\hat{b}_{\nu})$.\\
\indent The contact Hamiltonian $\hat{H}_{\alpha1}=\sum_{\alpha_1\in L,R}^{}\sum_{k\sigma'}^{}\epsilon_{\alpha_1 k\sigma'}\hat{c}_{\alpha_1k\sigma'}^{\dagger}\hat{c}_{\alpha_1k\sigma'}$ represents a reservoir of non-interacting and spin degenerate electrons with momentum eigenstate $k \sigma'$ and eigen energies $\epsilon_{\alpha_1k \sigma'}$. On the other hand, the reservoir Hamiltonian $\hat{H}_{\alpha_2}=\sum_{\alpha_2\in H,C }^{}\sum_{r}^{}\hbar\omega_{\alpha_2r}\hat{B}_{\alpha_2r}^{\dagger}\hat{B}_{\alpha_2r}$ characterizes a bath of independent phonon modes $r$ with frequencies $\omega_{\alpha_2r}$. Here  $\hat{c}_{\alpha_1k\sigma'}^{\dagger}(\hat{c}_{\alpha_1k\sigma'})$, creates (annihilates) an electron with momentum $k$ and spin $\sigma' \in \uparrow, \downarrow$ in the contacts $\alpha_1\in L,R$. Similarly $\hat{B}_{\alpha_2r}^{\dagger}(\hat{B}_{\alpha_2r})$ creates(annihilates) a phonon mode $r$ in the reservoirs $\alpha_2\in H,C$ of energies $\hbar\omega_{\alpha_2r}$. The spin independent coupling energy between the dot and the contact electrons is $\tau_{\alpha_1}^{el}$ and the mode independent coupling energy between the dot and the reservoir phonons is $\tau_{\alpha_2}^{ph}$. Electronic tunneling processes between the dot and the contacts and the phonon relaxation processes from the dot to the reservoirs are described by the Hamiltonian $\hat{H}_{T}=\sum_{k\sigma',\sigma}^{}[\tau_{\alpha_1}^{el}\hat{c}_{\alpha_1k\sigma'}^{\dagger}\hat{d}_{\sigma}+H.c.]  +\sum_{\nu,\alpha_2r}^{}\tau_{\alpha_2}^{ph}(\hat{B}_{\alpha_2r}^{\dagger}+\hat{B}_{\alpha_2r})(\hat{b}_{\nu}^{\dagger}+\hat{b}_{\nu})$, where $H.c.$ represents the Hermitian conjugate.\\
\indent The standard polaron transformation of the dot Hamiltonian is carried out which leads to the renormalization of the on-site energy ($\tilde{\epsilon}_{\sigma}=\epsilon_{\sigma}-\lambda_{\nu}^2\hbar\omega_{\nu}$), and the Coulomb interaction term ($\tilde{U}=U-2\lambda_{\nu}^2\hbar\omega_{\nu}$). The energy eigenvalues of the renormalized dot Hamiltonian become $E=\tilde{E}_{\sigma}+N\hbar\omega_{\nu}$, where  $\tilde{E}_{\sigma}={0,\tilde{\epsilon}_{\uparrow},\tilde{\epsilon}_{\downarrow}, \tilde{\epsilon}_{\uparrow}+\tilde{\epsilon}_{\downarrow}+\tilde{U}}$. This also leads to the renormalization of electron coupling energies $\tilde{\tau}_{\alpha_1}^{el}=\tau_{\alpha_1\sigma'\sigma}^{el}exp[-\lambda_{\nu}(\hat{b}_{\nu}-\hat{b}_{\nu}^{\dagger})]$. The modification of phonon coupling energy $\tau_{\alpha_2}^{ph}$ is neglected considering the weak coupling between the dot and the reservoir phonons. Using these, the rate of electron tunneling $\gamma_{\alpha_1}$ and phonon relaxation $\beta_{\alpha_2}$ are evaluated using the Fermi's golden rule: $\gamma_{\alpha_1}=\frac{2\pi}{\hbar}\sum_{\alpha_1}^{}\abs{\tilde{\tau}_{\alpha_1}^{el}}^2\rho_{\alpha_1\sigma}$ and
$\beta_{\alpha_2}=\frac{2\pi}{\hbar}\abs{\tau_{\alpha_2}^{ph}}^2D_{\alpha_2}$. Here $\rho_{\alpha_1\sigma}$ and $D_{\alpha_2}$ are the constant electron and phonon density of states associated with electrodes $\alpha_1$ and reservoirs $\alpha_2$.\\
\indent We use the approximation $\hbar\gamma_{\alpha_1} >> \hbar\beta_{\alpha_2}$, so that we can ignore the system damping \cite{Braig2003} and assume that each reservoir generates phonon currents independent of other reservoirs \cite{segal2006,de}.  We also set $\hbar\omega_{\nu}>>\hbar\gamma_{\alpha_1}$, so that any overlap between two adjacent phonon sidebands is excluded and subsequently, the electron tunneling events are also completely uncorrelated \cite{de,Theses,Brouw}.  The sequential tunneling limit is assumed \cite{Bli3,Timm} such that $k_BT>>\hbar\gamma_{\alpha_1},\hbar\beta_{\alpha_2}$ \cite{Leijnse2010}, where, charge and heat transport are described via the master equation framework by evaluating the steady state probability $P_{(n,q)}$ of an electron-phonon many-body state $(n,q)$, with $n$ electrons and $q$ phonons \cite{Beenakker}. Using $P_{(n,q)}$,  we can calculate the charge and heat currents associated with the electrodes $\alpha_1\in L,R$ and the reservoirs $\alpha_2\in H,C$\cite{Leijnse2010,de,Muralidharan2012,segal2006} as
	\begin{equation}
	\begin{gathered}
	I_{\alpha_1}=\sum_{n,q}^{}\sum_{n,q'}^{}-q\bigg[R_{(n',q')\rightarrow(n,q)}^{el_{\alpha_1}}P_{(n',q')}\\ -R_{(n,q)\rightarrow(n',q')}^{el_{\alpha_1}}P_{(n',q')}\bigg]\delta(n\pm1,n')
	\end{gathered}
	\label{Eq1}
	\end{equation}
	\begin{equation}
	\begin{gathered}
	I_{ph_{\alpha_2}}^{Q}=\sum_{n,q}^{}\sum_{n,q'}^{}\hbar\omega_{\nu}\bigg[R_{(n,q)\rightarrow(n',q')}^{ph_{\alpha_2}}P_{(n,q)}\\-R_{(n',q')\rightarrow(n,q)}^{ph_{\alpha_2}}P_{(n',q')}\bigg]\delta(n,n')\delta(q\pm1,q').
	\end{gathered}
	\label{Eq2}
	\end{equation}
	\indent The phonon relaxation rate $R_{(n,q)\rightarrow(n',q')}^{ph_{\alpha_2}}$  and electron tunneling rate $R_{(n,q)\rightarrow(n',q')}^{el_{\alpha_1}}$ between the two states $(n,q)$ and $(n',q')$ depend upon the Bose-Einstein and Fermi-Dirac function of the energy difference of the two states. They are also proportional to $\beta_{\alpha_2}$ and $\gamma_{\alpha_1}$. However in the presence of electron-phonon interaction, the effective electron tunneling rate $\gamma_{\alpha_1}^{eff}$ between the two states $(n,q)$ and $(n\pm1,q')$ becomes a function the \textit{Frank-Condon} \cite{Leturcq2008,Kochh2006} overlap factor($FC_{q,q'}$) between them such that
	\begin{equation}
	\begin{gathered}
	\gamma_{\alpha_1}^{eff}=\gamma_{\alpha_1}\abs{FC_{q,q'}}^2\\
	\gamma_{\alpha_1}^{eff}=\gamma_{\alpha_1}exp(-\lambda^2)\frac{k!}{K!}\lambda^{2(K-k)}[L_{k}^{K-k}(\lambda^2)]^2,
	\end{gathered}
	\label{Eq3}
	\end{equation}
where $L_{k}^{K-k}$ is the associated Laguerre polynomial with $k=min(q,q')$ and $K=max(q,q')$. When $q\neq q'$ the charge transport leads to phonon generation (or absorption) in the dot with a rate $GE_{ph}^{\alpha_1}=\sum_{n,q}^{}\sum_{n\pm 1,q'}^{}(q'-q)P_{(n,q)}R_{(n,q)\rightarrow(n\pm 1,p')}^{\alpha_1}$. One should note from \eqref{Eq3}, that when $\lambda=0$, $\gamma_{\alpha_1}^{eff}$ becomes zero unless $q=q'$. In that case $GE_{ph}^{\alpha_1}$ vanishes and it implies that there is no charge assisted phonon generation in the absence of electron-phonon interaction. The generated phonons are further removed by the reservoirs with a rate of $RE_{ph}^{\alpha_2}$ given by \cite{Siddiqui2006}:
	\begin{equation}
	\begin{gathered}
	RE_{ph}^{\alpha_2}=\beta_{\alpha_2}\frac{\langle N_{ph}\rangle-N_{ph}^{eq}}{1+N_{ph}^{eq}}.
	\end{gathered}
	\label{Eq4}
	\end{equation}
	In the above equation, $N_{ph}^{eq}$ is the average equilibrium phonon number. The average phonon occupation is defined as $\langle N_{ph}\rangle=\sum_{n,q}^{}qP_{(n,q)}$. If $GE_{ph}^{\alpha_1}$ exceeds $RE_{ph}^{\alpha_2}$, phonons accumulate in the dot and the dot temperature deviates from the reservoir temperature. The expression for the dot temperature $T_M$, can be derived from the Boltzmann ratio with a quasi-equilibrium approximation \cite{oppen1} as:
	\begin{equation}
	T_{M}=\frac{\hbar\omega_{\nu}}{k_B}\bigg[ln\bigg(\frac{P_{n,q}}{P_{n,q+1}}\bigg)\bigg]^{-1}
	\label{Eq5}
	\end{equation}
	We now denote $I_{\alpha_1}$, $I_{ph_{\alpha_2}}^{Q}$, $GE_{ph}^{\alpha_1}$, $RE_{ph}^{\alpha_2}$ as $I$, $I_Q^{ph}$, $GE_{ph}$, $RE_{ph}$, respectively, to maintain notational simplicity.\\
	\begin{figure}[]
		\begin{center}
			\subfigure[]{\includegraphics[width=0.22\textwidth, height=0.22\textwidth]{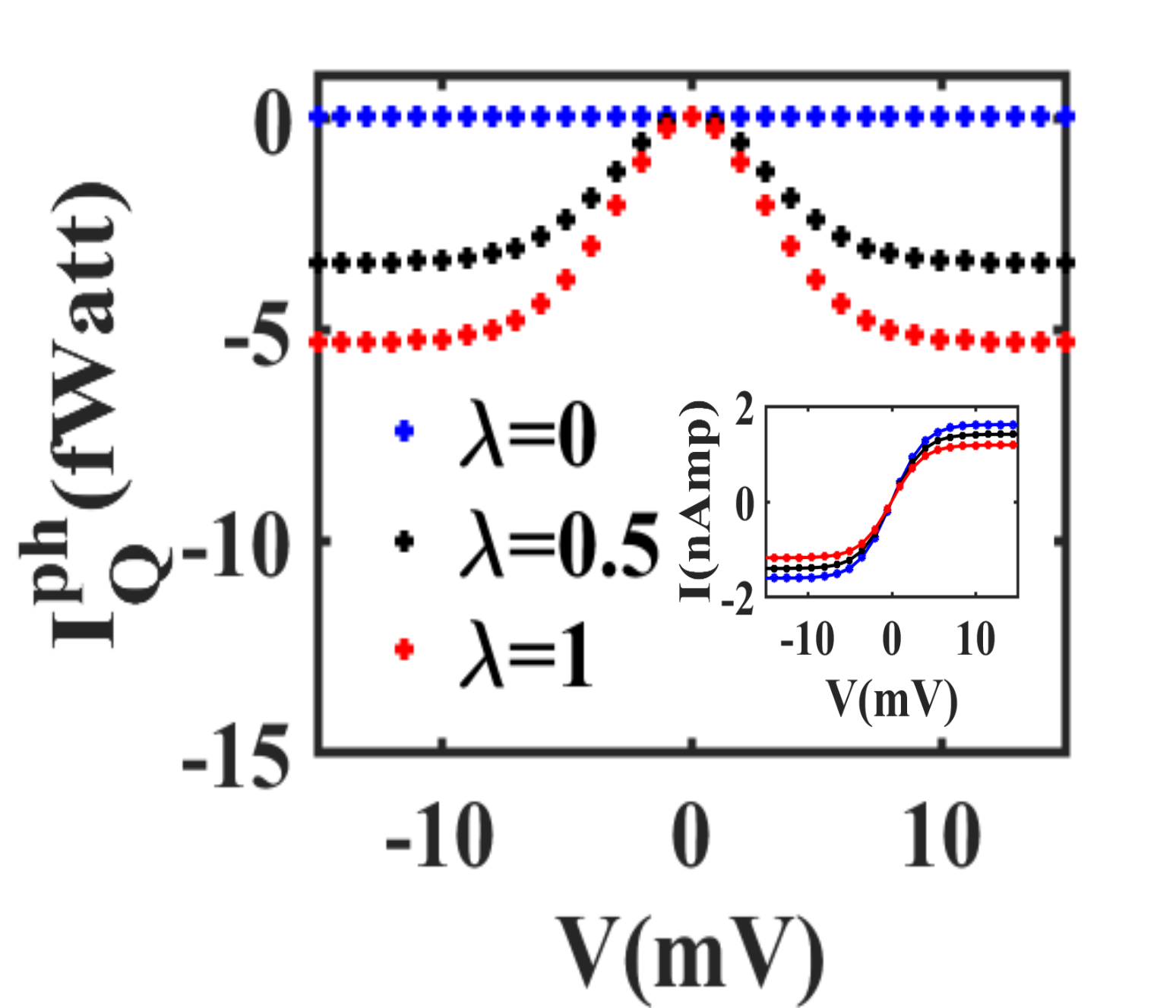}\label{2a}}
			\quad
			\subfigure[]{\includegraphics[width=0.22\textwidth, height=0.22\textwidth]{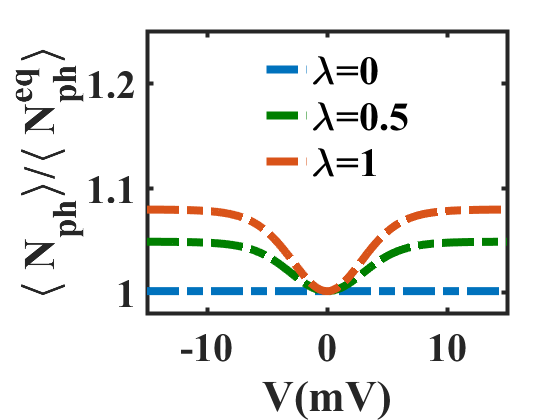}\label{2b}}
			\quad
			\subfigure[]{\includegraphics[width=0.22\textwidth, height=0.22\textwidth]{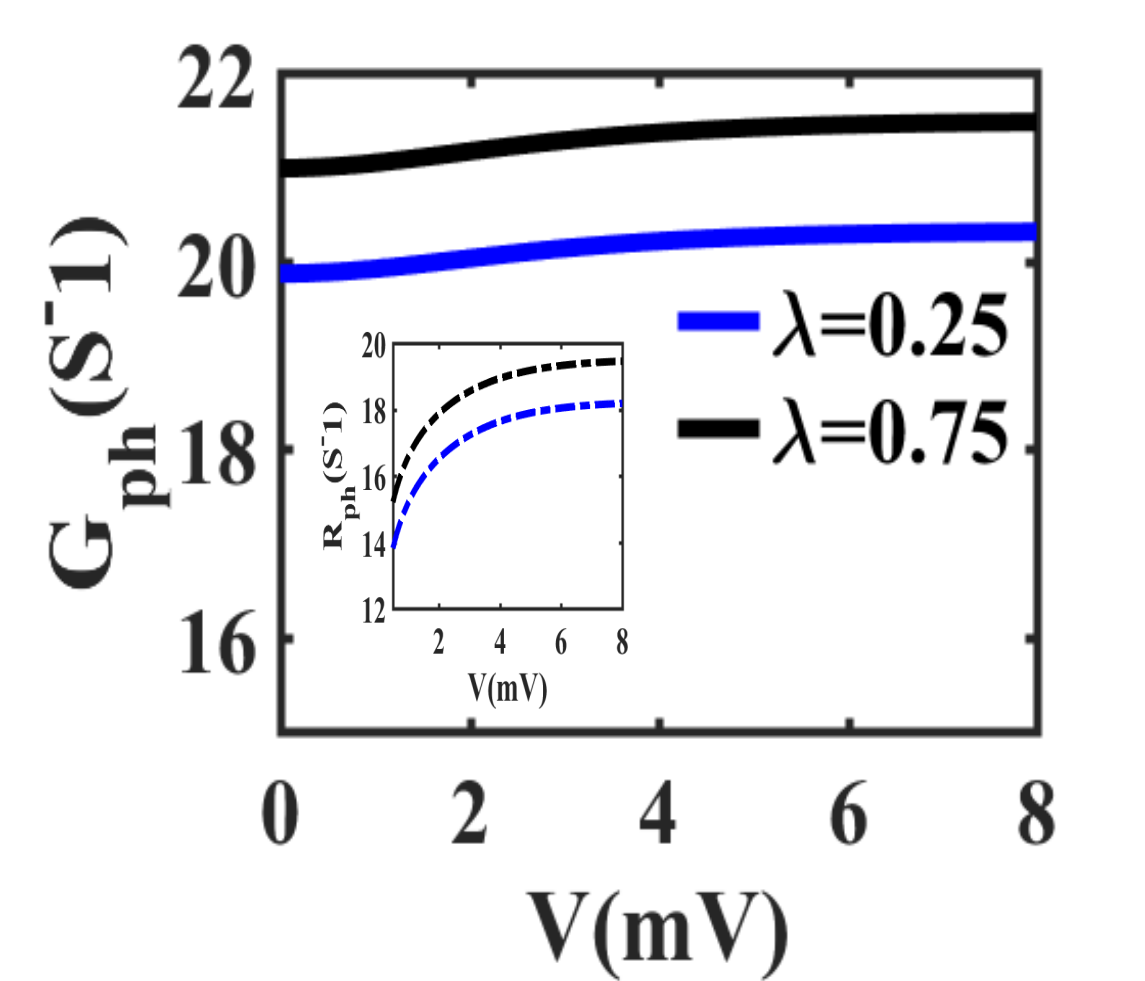}\label{2c}}
			\quad
			\subfigure[]{\includegraphics[width=0.22\textwidth, height=0.22\textwidth]{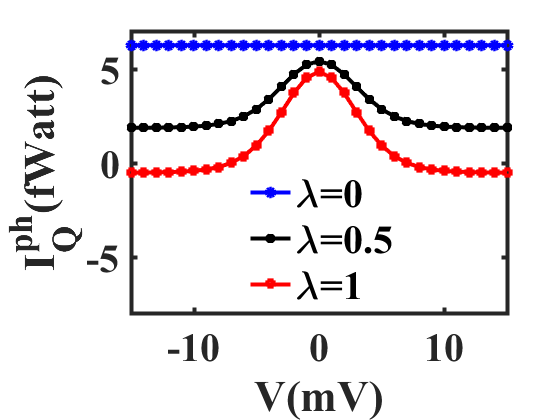}\label{2d}}
			
		\end{center}
		\caption{Phonon transport due to charge induced phonon generation: (a) Variation of $I_Q^{ph}$ with voltage for different $\lambda$ keeping $\Delta T_{ph}=0$. For high bias, $I_Q^{ph}$ saturates along with the charge current as shown in the inset. (b) Deviation of $\langle N_{ph}\rangle$ from $N_{ph}^{eq}$ as voltage and $\lambda$ are varied. (c) We note the dominance of the charge assisted phonon generation rate over the reservoir controlled phonon extraction rate leads to an increase in $\langle N_{ph}\rangle$ over $\langle N_{ph}^{eq}\rangle$. (d) Variation of $I_Q^{ph}$ with voltage when $\Delta T_{ph}$ is non-zero.}
	\end{figure}
\indent{\it{Results.}} The principal signature of charge assisted non-equilibrium phonon generation is the variation of $I_Q^{ph}$ with voltage bias. In Fig.~\ref{2a}, we note that when $\lambda=0$, the phonon heat current $I_Q^{ph}$ vanishes for all voltages, provided there is no thermal gradient. It is consistent with the fact that in the absence of electron-phonon interaction, charge assisted phonon generation does not occur and hence, the dot phonons are in equilibrium with the reservoir phonons. The scenario changes when we turn on a finite electron-phonon interaction ($\lambda \neq 0$), where a non-trivial voltage dependence of $I_Q^{ph}$ results, as seen in Fig.~\ref{2a}. However for large voltages, $I_Q^{ph}$ levels off since the charge current becomes constant, as shown in the inset of Fig.~\ref{2a}. 
\begin{figure}[!htb]
		\begin{center}
			\subfigure[]{\includegraphics[width=0.22\textwidth, height=0.20\textwidth]{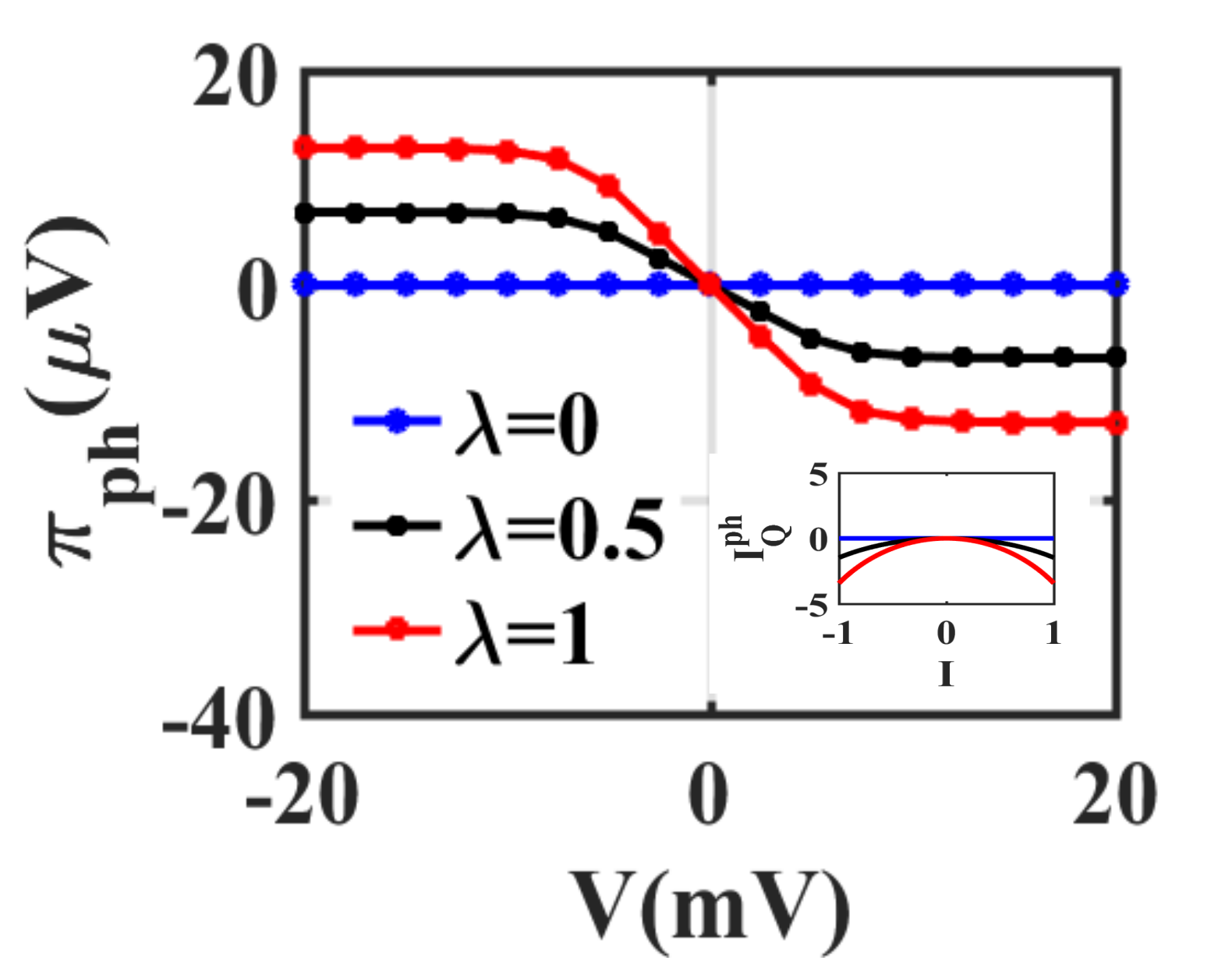}\label{3a}}
			\quad
			\subfigure[]{\includegraphics[width=0.22\textwidth, height=0.20\textwidth]{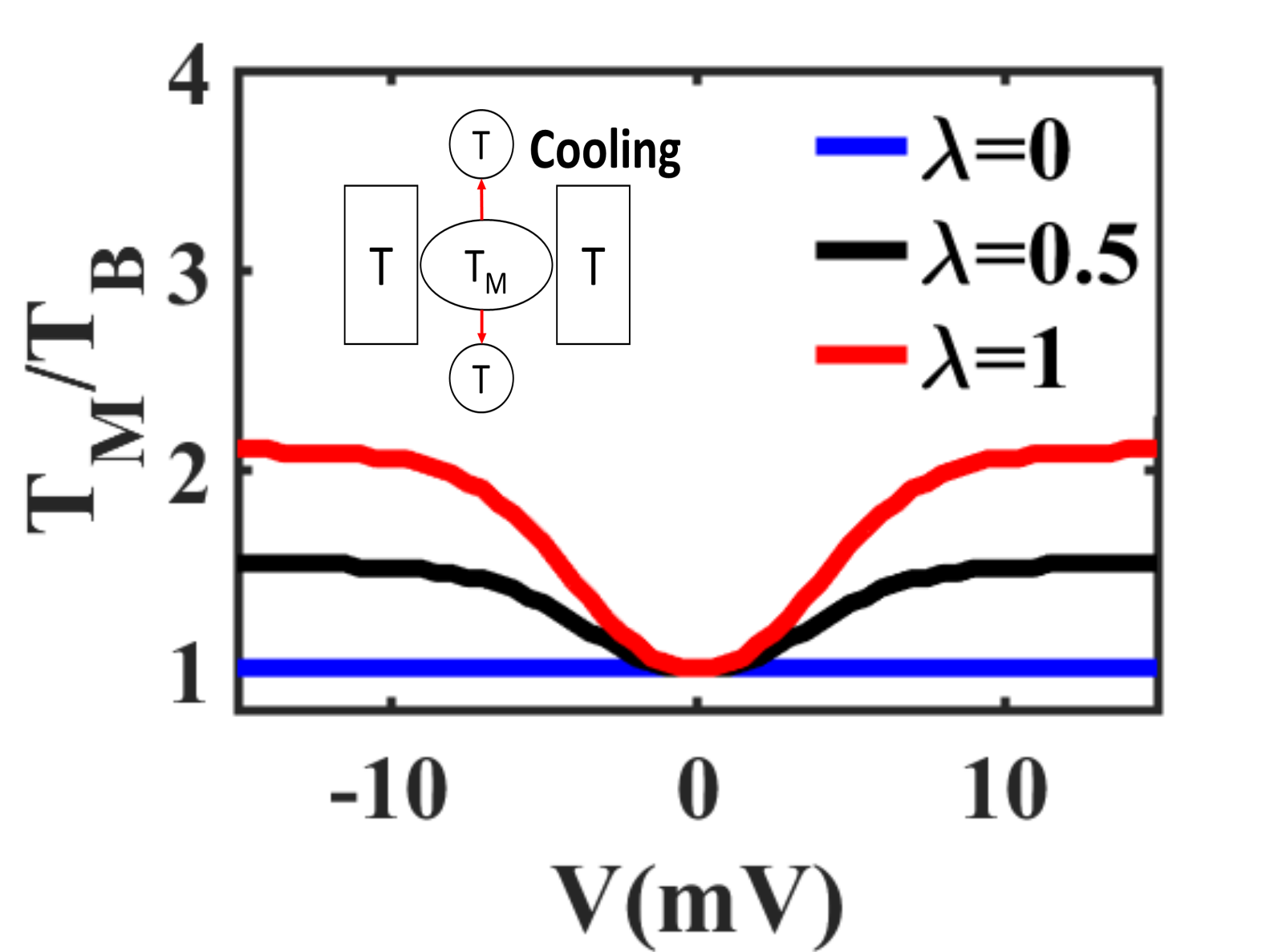}\label{3b}}
			\end{center}
		\caption{Non-linear phonon pelter effect.(a) plots the variation of $\pi_{ph}$ with voltage for different $\lambda$.The inset shows the variation of $I_Q^{ph}$ with the charge current $I$. $\pi_{ph}$ is extracted from the gradient of the inset plot.(b) shows how a  thermal gradient is created between the dot and reservoirs as an outcome of phonon Peltier.}
	\end{figure}
As explained earlier, a finite $\lambda$ facilitates phonon generation and the average dot phonon distribution $\langle N_{ph}\rangle$ deviates from $N_{ph}^{eq}$ as noted in Fig.~\ref{2b}. Hence, as dictated by \eqref{Eq4}, the phonon removal rate changes with voltage and ultimately leading to the voltage dependence of $I_Q^{ph}$. Throughout this work we maintain $\gamma_{L(R)}>\beta_{H(C)}$, resulting in $GE_{ph}$ exceeding $RE_{ph}$ as depicted in Fig.~\ref{2c}. It leads to the phonon accumulation inside the dot and $\langle N_{ph}\rangle$ always exceeding $\langle N_{ph}^{eq}\rangle$. As a result, $I_Q^{ph}$ remains negative, i.e.,  $I_Q^{ph}$ flows from the dot to the reservoirs.\\
\indent An important point to be noted is that the variation of $I_Q^{ph}$ takes place only in the non-linear voltage range, i.e., when $qV_{app}>>k_{B}T$. In Fig.~\ref{2d}, we plot the variation of $I_Q^{ph}$ with applied bias when $\Delta T_{ph}\neq0$, where we note the curves being vertically offset when compared with Fig.~\ref{2a}. The voltage dependence of $I_Q^{ph}$ immediately hints at a novel Peltier effect which can be thought of as the \textit{Phonon Peltier effect} rather than the conventional \textit{electronic Peltier effect}.\\
\indent The phonon Peltier coefficient is calculated as $\pi_{ph}=I_Q^{ph}/I$ in analogy with electronic Peltier coefficient $\pi_{el}=I_Q^{el}/I$, where $I_Q^{el}$ is the electronic heat current, provided there is no temperature gradient (i.e., $T_H=T_C=T_B$). The variation of $\pi_{ph}$ is plotted as a function of voltage in Fig.~\ref{3a}. In our simulations, we plot $\pi_{ph}=dI_Q^{ph}/dI$ to avoid the discontinuity at the short-circuit ($V=0$) point and $\pi_{ph}$ is extracted from the gradient of the $I_Q^{ph}-I$ plot shown in the inset of Fig.~\ref{3a}. It shows that for non-zero $\lambda$, $\pi_{ph}$ increases almost linearly within a small voltage range and levels off at large voltages as the charge current saturates. This is in contrast with electronic Peltier coefficient which always changes with voltage because electronic heat current directly depends on voltage bias. On the other hand, $I_Q^{ph}$ primarily depends on $\langle N_{ph}^{eq}\rangle$ which saturates at large bias. \\
\begin{figure}[!htb]
		\begin{center}
			\subfigure[]{\includegraphics[width=0.22\textwidth, height=0.20\textwidth]{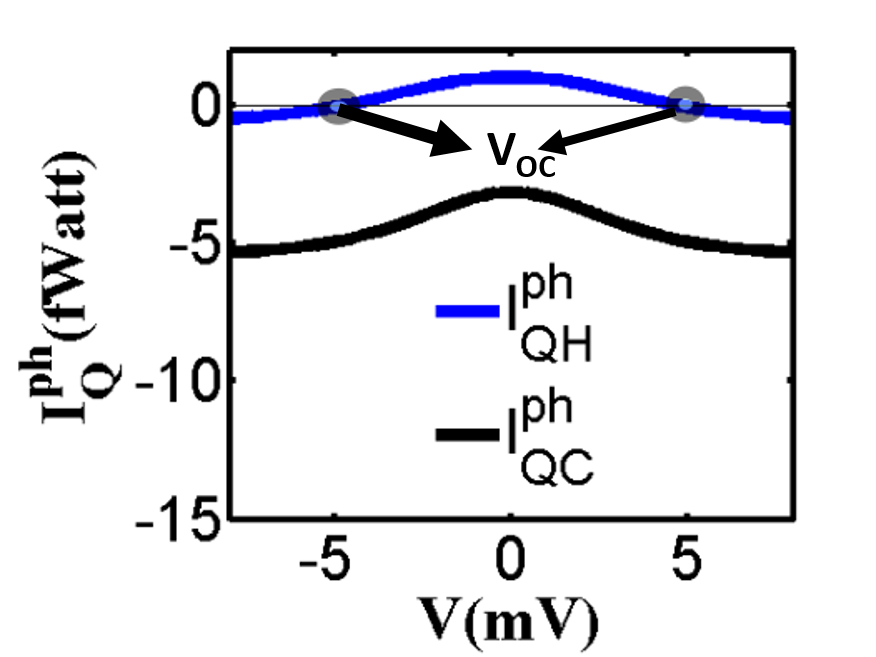}\label{4a}}
			\quad
			\subfigure[]{\includegraphics[width=0.22\textwidth, height=0.20\textwidth]{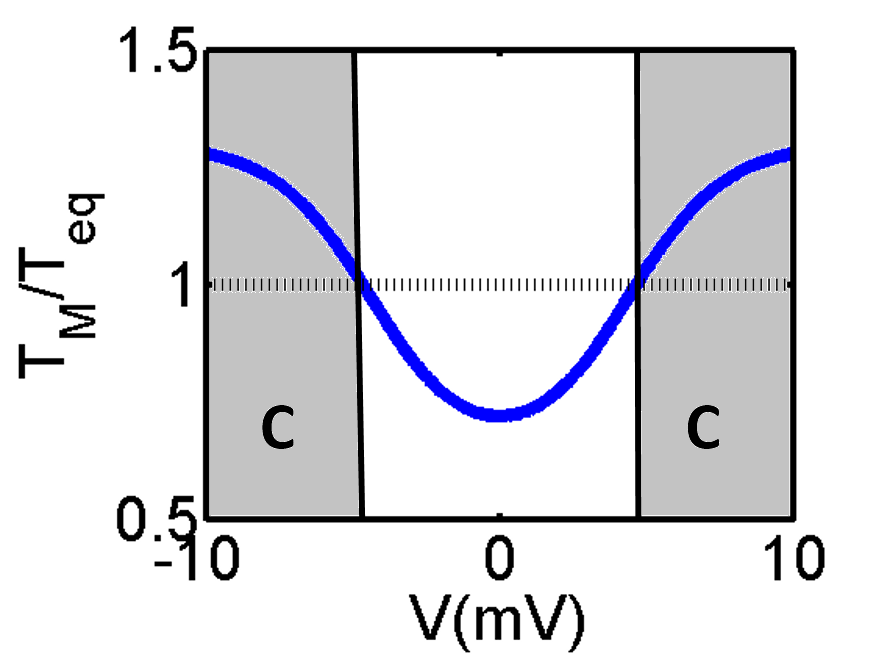}\label{4b}}
			\quad
			\subfigure[]{\includegraphics[width=0.22\textwidth, height=0.20\textwidth]{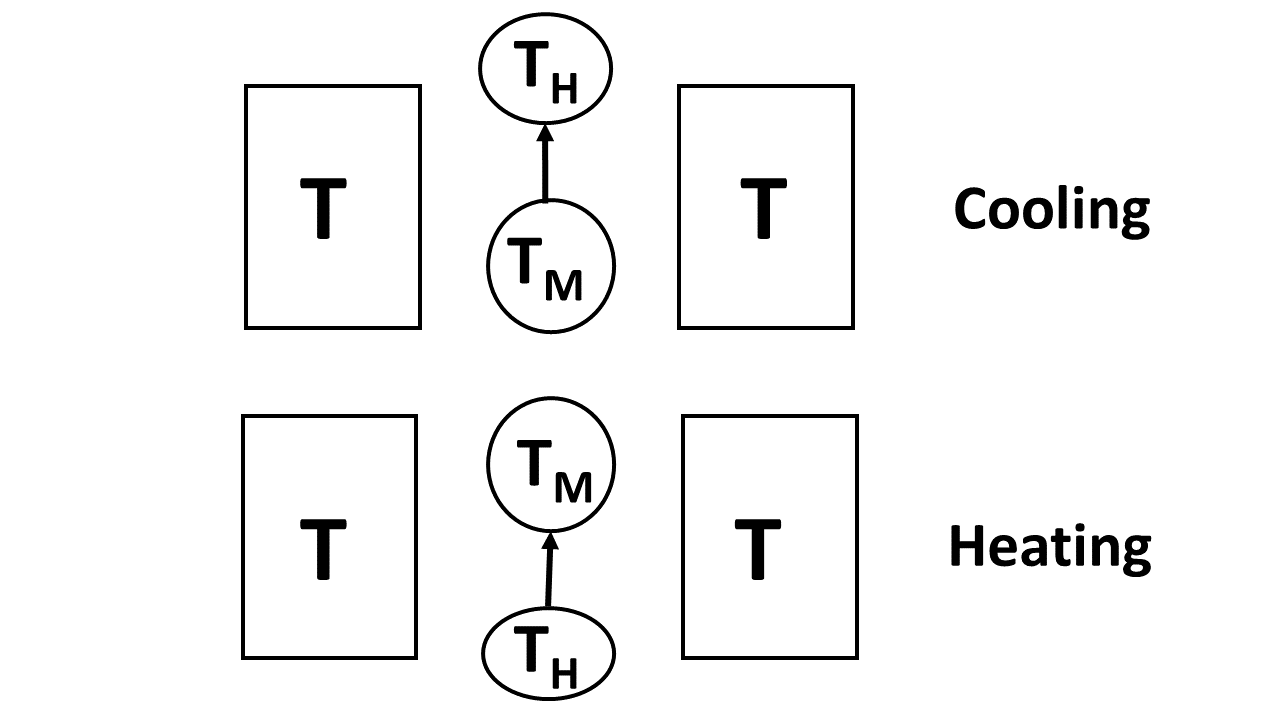}\label{4c}}
			\quad
			\subfigure[]{\includegraphics[width=0.22\textwidth, height=0.20\textwidth]{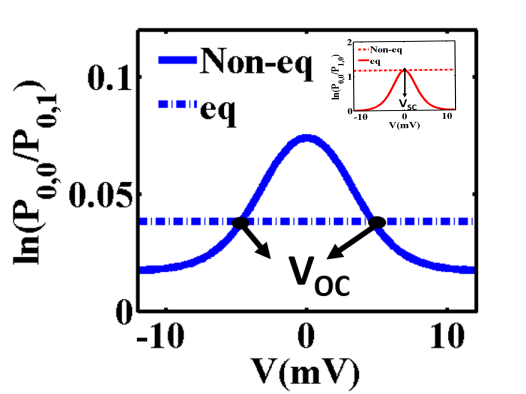}\label{4d}}
		\end{center}
		\caption{Peltier cooling of the dot. (a) Variation of phonon heat current as a function of voltage for the Hot and the Cold bath. For the Hot bath, at $V=\pm V_{OC}$, $I_Q^{ph}$ reverses its direction. The dot is cooled via the Hot bath when $V>V_{OC}$. (b) Variation of $T_M$ (normalized by the reservoir temperature) with respect to voltage for the Hot reservoir case. (c) Schematic of heating and cooling by the Hot reservoir. (d) We note that phonons are brought to the equilibrium distribution by charge transport at voltages $V=\pm V_{OC}$ in the Hot reservoir case. The inset points at the electronic equilibrium point which is at $V=0$. }
	\end{figure}
\indent Typically, the Peltier effect causes a temperature gradient when subject to an electrical excitation. In this case, we note a temperature difference is created between the dot and the reservoir phonons, which we estimate via the dot temperature $T_M$ using a thermometer probe \cite{de}. Figure ~\ref{3b} shows a clear deviation of $T_M$ from the reservoir temperature $T_B$. Since $\gamma_{\alpha_1}>\beta_{\alpha_2}$, phonons accumulate in the dot and hence $T_M$ is always greater than $T_{B}$. Consequently, the phonon current will flow away from the dot as shown in the inset of Fig.~\ref{3b} and clearly signifies Peltier cooling.\\
\indent So far we have the electrode temperature equal to the reservoir temperature, i.e.,  $T_{L(R)}=T_{H(C)}$. Let us now consider two cases: (a) the hot reservoir,  $T_{H(C)}>T_{L(R)}$ and (b) the cold reservoir $T_{H(C)}<T_{L(R)}$. The phonon heat currents now renamed $I_{QH}^{ph}$ and $I_{QC}^{ph}$ are plotted in Fig. ~\ref{4a} as a function of voltage. We notice that $I_{QH}^{ph}$ vanishes at a voltage of $\pm V_{OC}$. Interestingly, when $\abs{V}>V_{OC}$, the hot reservoir cools the dot instead of heating it. This sort of counterintuitive cooling is an important consequence of electron-phonon interaction. In hot reservoir case, at $\pm V_{OC}$,  $T_M$ equals the reservoir temperature as noted in Fig.~\ref{4b}. However, no such direction reversal of  $I_{QC}^{ph}$  is noticed in the cold reservoir case. The block diagram in Fig.~\ref{4c} depicts the cooling and heating schematic in the hot reservoir case. \\ 
\begin{figure}[!htb]
		\begin{center}
			\subfigure[]{\includegraphics[width=0.22\textwidth, height=0.22\textwidth]{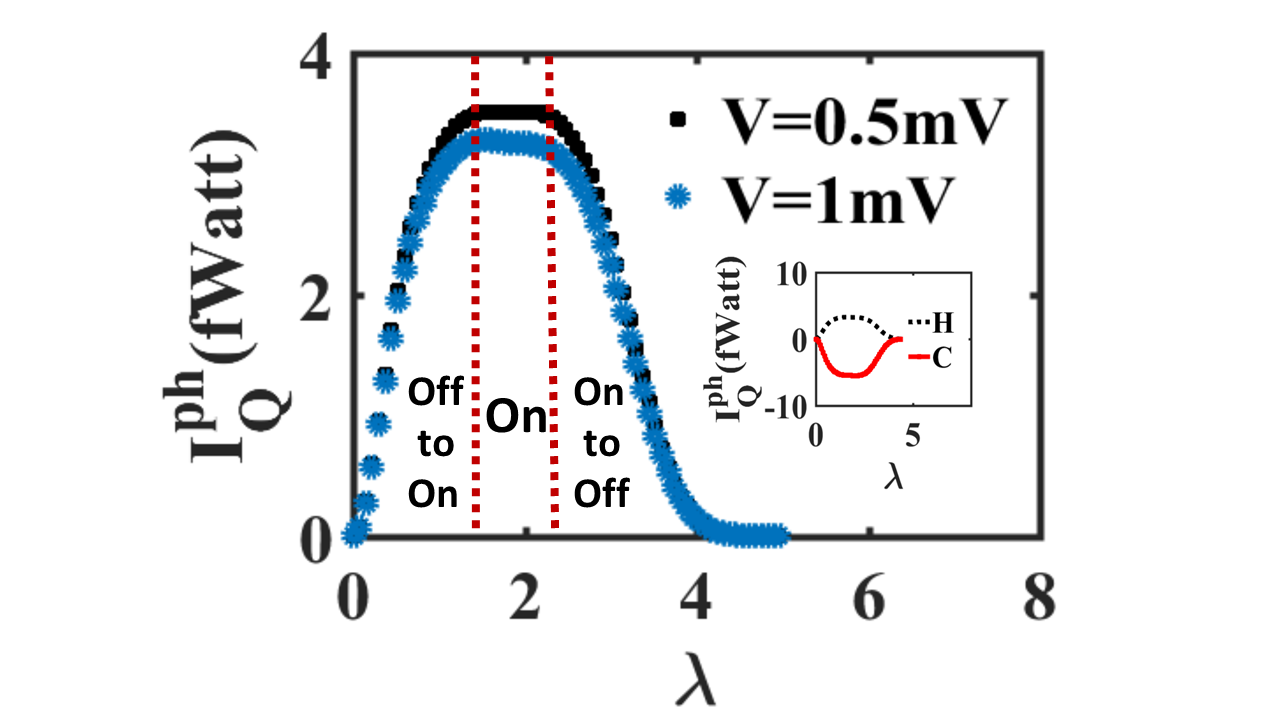}\label{5a}}
			\quad
			\subfigure[]{\includegraphics[width=0.22\textwidth, height=0.22\textwidth]{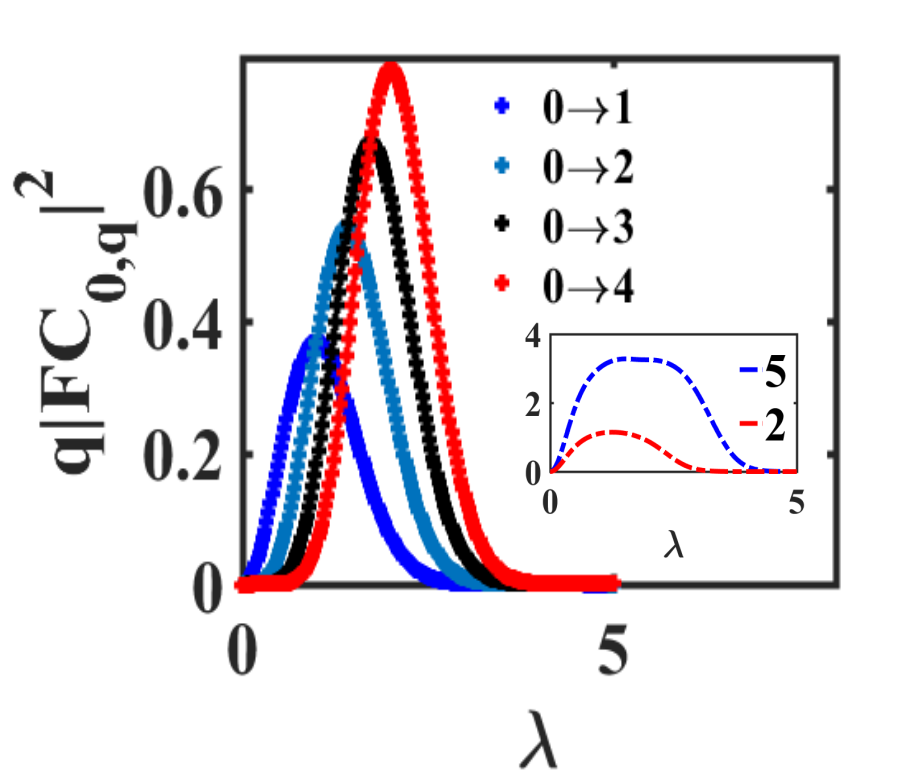}\label{5b}}
		\end{center}
		\caption{Phonon switch facilitated by $\lambda$: (a) Variation of $I_Q^{ph}$ with $\lambda$ with varying voltage. The profile gives an idea of the phonon switch with different operating regions. The inset shows how the polarity of $I_Q^{ph}$ reverses for the Hot ($T_B>>T_C$) and the Cold($T_B>>T_C$) reservoir.(b) Variations of different phonon generation processes with $\lambda$ that control the switch operation. The inset plot shows how more number of phonon sidebands aid efficient switching.}
	\end{figure}
\indent Generally, in the absence of electron-phonon interaction and thermal bias, both electrons and phonons equilibrate at the short-circuit point $V=0$. As the electron-phonon interaction $\lambda$ is turned on, the phonon equilibrium point is modulated by the charge flow and is directly influenced by the reservoir temperature. In the hot reservoir case,  the phonon equilibrium point is reached at $\pm V_{OC}$ as plotted in Fig.~\ref{4d}, whereas in the cold reservoir case, phonons are always out of equilibrium. The modulation of phonon equilibrium by the electrons is an interesting outcome that results in the counterintuitive cooling that we have noted.\\
\indent We note that if one wishes to switch phonon currents, voltage is not a suitable control parameter because $I_Q^{ph}$ cannot even turn on when $\lambda=0$. Even in the finite $\lambda$ case, $I_Q^{ph}$ can be turned on but not turned off. However, switching of phonon currents can be achieved by varying $\lambda$ at a constant voltage. In Fig.~\ref{5a}, we investigate the variation of $I_Q^{ph}$ as a function of $\lambda$ at fixed voltage. The phonon current first shoots up, levels off and finally falls as $\lambda$ is ramped up. Therefore, $I_Q^{ph}$ can be switched on and off via the modulation of $\lambda$. According to \eqref{Eq3}, the phonon generation rate associated with the charge transfer between two phonon states with phonon number $q_1$ and $q_2$ is proportional to $(q_2-q_1)\abs{FC_{q_1,q_2}}^2$. In Fig.~\ref{5b}, we plot the variation of $\abs{FC_{q_1,q_2}}^2$ as a function of $\lambda$ for different values of $q_2$ with $q_1=0$. It is observed that for the small values of $\lambda$, $\abs{FC_{q_1,q_2}}^2$  shoots up and the switch is turned ON as we increase $\lambda$. As $\lambda$ is increased further, $\abs{FC_{q_1,q_2}}^2$ starts to decay. The decay starts at earlier $\lambda$ for smaller values of $q_2$ and shifts to larger values of $\lambda$ with increasing $q_2$. At intermediate $\lambda$, $I_Q^{ph}$  levels off and the switch remains ON. At large $\lambda$ all the processes fall sharply, and the switch is turned OFF. Recently, it was demonstrated \cite{Ilani} that $\lambda$ can be tailored in a suspended CNT-quantum dot, by applying a gate voltage. Our simulation proposes that this kind of electron-induced thermal switching might be accomplished in such kinds of structures.\\
\indent Phonon computation is an emerging area that aims to convert waste heat (phonon currents in our case) to electrical signals which can be re-implemented in logic design. A major challenge in that course is to modulate the phonon current by electrical means. Our proposal for phonon switching can accomplish such a goal. This switching can be used to design thermal pulse generators and thermal multi-vibrators. For efficient switching, the 'ON' region should be more broadened, and that can be achieved by increasing the number of phonon sidebands in the voltage window as shown in the inset of Fig.~\ref{5b}.\\
\indent The non-linear transport regime in a quantum dot heat engine described by Anderson-Holstein model was investigated in detail. it was shown that a finite electron-phonon interaction leads to a charge induced phonon generation that stimulates a phonon current even in the absence of a thermal gradient. This gave rise to  the non-linear phonon Peltier effect which increases with the electron-phonon interaction. Utilizing the reversal of phonon currents via charge induced phonon accumulation, we demonstrated that the heat engine can be cooled through a hot bath. In further exploring possibilities that can arise from this effect, we proposed a charge-induced phonon switching mechanism as a building block for phonon computation.
\bibliographystyle{apsrev}
\bibliography{refrences}
\end{document}